\documentclass[a4paper]{VisionStyle}
\usepackage{epsfig}

\begin{document}

\title{XMM -Newton observations of merging clusters of galaxies: A3921 and A1750}

\author{E.\,Belsole\inst{1} \and  J-L.\,Sauvageot\inst{1} \and 
  R. \,Teyssier\inst{1} } 

\institute{Service d'Astrophysique, DSM/DAPNIA/CE-Saclay, bat 709 l'Orme des Merisiers, 919191 Gif sur Yvette Cedex, France }
\maketitle 

\begin{abstract}
\end{abstract}
We show the XMM-Newton guaranted time observations of 2 clusters of galaxies in a different stage of merger.
%New Visions of the X-ray Universe in the XMM-Newton and Chandra era
%ESTEC - Noordwijk   26-30 November 2001

\section{Introduction}
In the hierachical scenario of structure formation in the Universe, clusters are assembled by mergers of smaller systems. Observations and simulations show that merging strongly changes the physical characteristics of the intracluster medium. The study of this dynamical process would give us new insights  into the physical process involved and better constraints in a more general cosmological context. A good indication of merging events is given by the detection of substructures in a cluster because these are destroyed in a few Gyr. Our XMM-Newton guaranteed time programme allows us to trace the evolution of the intracluster medium, and the cluster itself, by observing clusters in different stages of the merger process. We will present the cluster A3921 which seems to be in a stage of  ongoing merging. The EPIC observations show an increase of the temperature toward the center and a compression region between the main cluster and the infalling subcluster. We will show how AMR simulations of a galaxy cluster of the same type as  A3921 predict this behaviour. The recent observation of our second  guaranteed time target, A1750, allows us to show an earlier merging event between two subclusters of roughly equal mass. The EPIC/MOS data indicate the presence of a shocked region between the two subunits even in a so early stage of merger.

\section{A3921}

 A3921  has been observed by ROSAT and GINGA revealing that a substructure is falling onto the  main relaxed cluster from behind with a velocity of ~ 1000 km s$^{-1}$. The substructure is located at a projected distance of 1.2 Mpc (z=0.094; h$_o$= 0.5)   and it has a mass   ~ 1/3  of the main component. A3921 was observed by  XMM-Newton  on October 2000 for a total exposure of 30 ks. The analysis of EPIC/EMOS  data had already shown a complex structure in the center of the main cluster which seems highly perturbed (Belsole et al. 2001) and a filamentary structure in the hardness ratio map in the region between the two components which has  an higher temperature than the main cluster (Sauvageot et al. 2001).
\begin{figure}[th]
  \begin{center}
    \epsfig{file=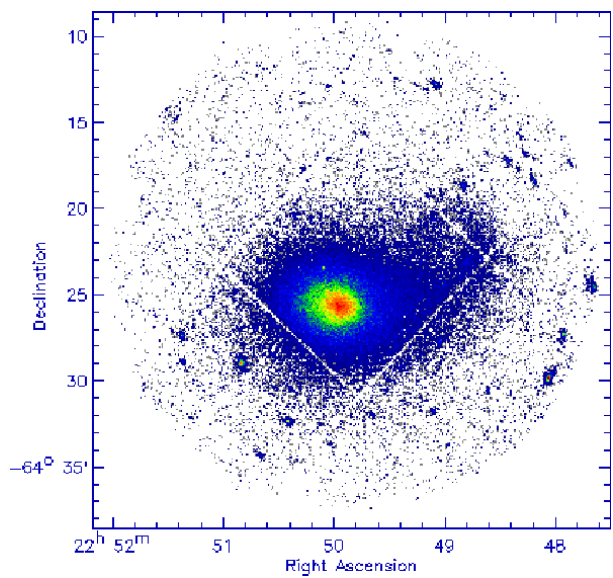, width=8cm}
  \end{center}
\caption{A summed EMOS1 \& EMOS2 image in the energy band 0.5-5.0 keV. North is up, east to the left}  
\label{fauthor-E1_fig:fig1}
\end{figure}

\subsection{Imaging}
The surface brightness profile is obtained excluding point sources, the falling cluster and the region q$<$1.25' corresponding to the central excess.
The central radius and b parameter are in good agreement with ROSAT/PSPC results

\begin{figure}[th]
  \begin{center}
    \epsfig{file=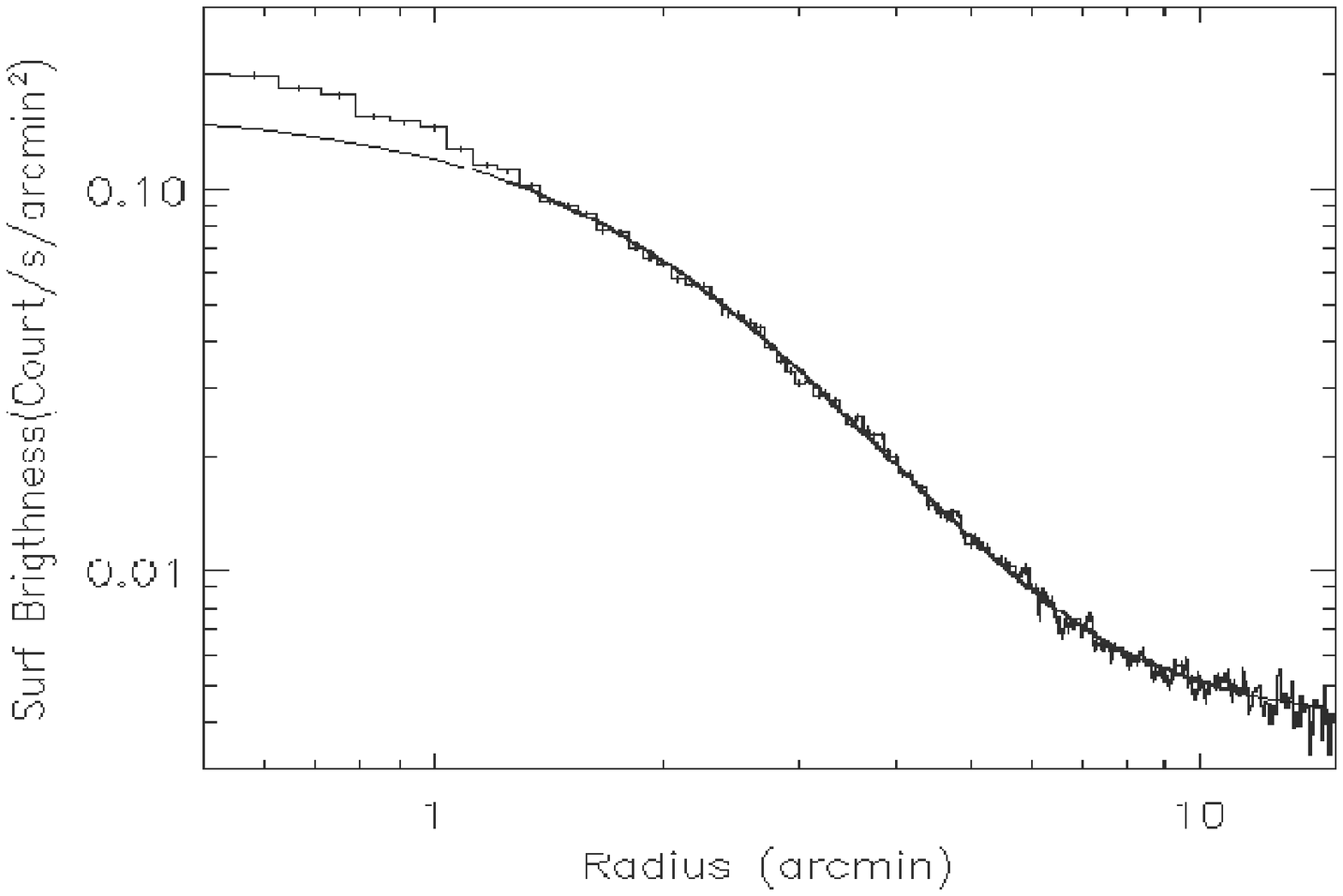, width=7cm}
    \epsfig{file=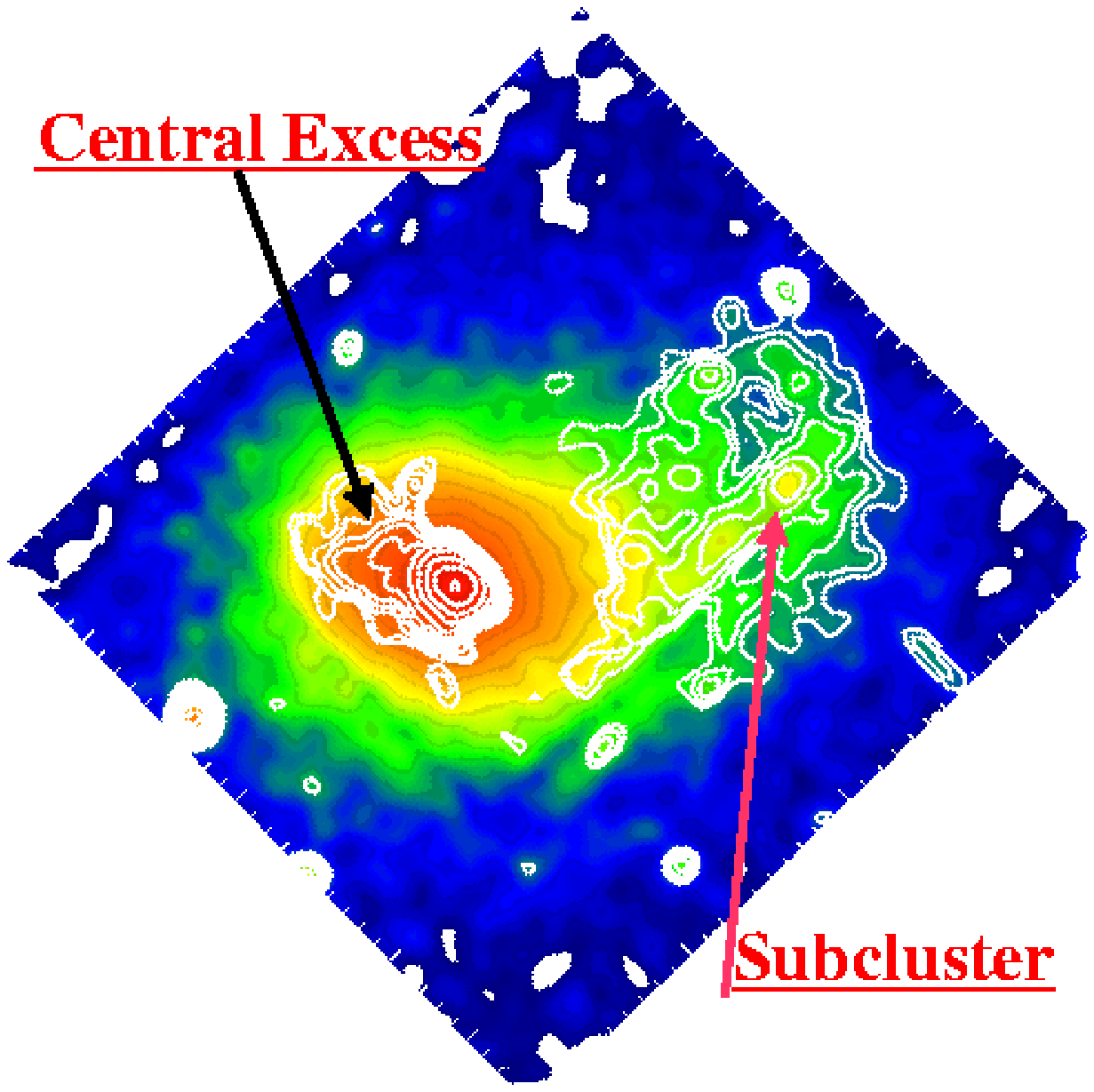, width=7cm}
  \end{center}
\caption{{\em Top:} A3921 surface brightness profile. {\em Bottom:} Wavelet image of A3921 subtracted of a the 2D beta model. Contours are the central enanchement and the subgroup structure.}  
\label{fauthor-E1_fig:fig2}
\end{figure}

\subsection{Spectro-imaging}
The temperature map is obtained  by a combined fit (absorbed mekal model in XSPEC) of the EMOS1 and EMOS2 spectra in each box, after background subtraction and vignetting correction. 
The temperature at the center of the main component is kT = 5.3 where the  surroundings have a slightly lower temperatures, thus the cooling flow hypothesis is  ruled out. The region between the two merging clusters is at a mean temperature of 5.8 keV. Curiously the small subcluster has a relatively high temperature (5.8 keV).

\begin{figure}[th]
  \begin{center}
    \epsfig{file=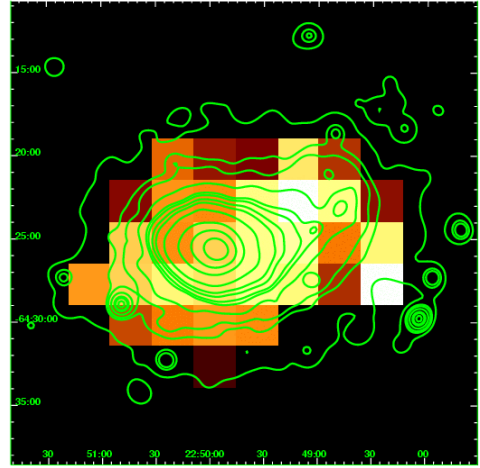, width=8cm}
  \end{center}
\caption{Temperature map. Contours correspond to the image in figure 1 after adaptive smoothing}  
\label{fig3:fig3}
\end{figure}

\section{AMR  simulations}

We show an example of a merging cluster extracted from a high-resolution cosmological simulation, using the newly developed Adaptive Mesh Refinement code RAMSES 
(Teyssier, 2001). The two color maps show the bolometric X-ray luminosity and the luminosity-weighted temperature. This simulated cluster is of the same type of A3921 in order directly to indicate comparisons between observations and simulations. The hot compression wave in between the two interacting clusters is clearly visible in the figure. For more detalis about AMR simulations,  see the poster by J-L. Sauvageot et al. at this conference.
\begin{figure}[ht]
  \begin{center}
    \epsfig{file=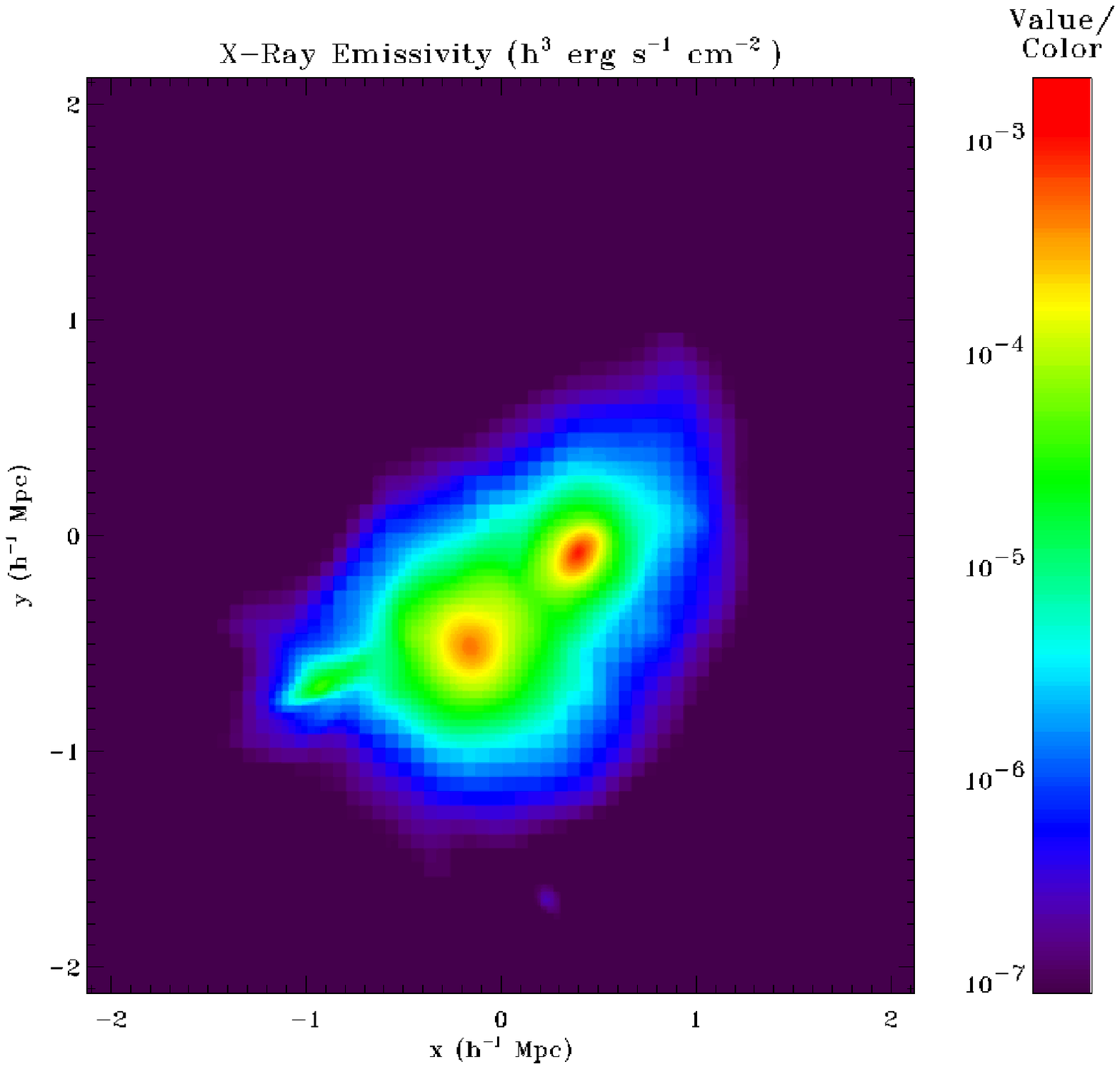, width=8cm}
    \epsfig{file=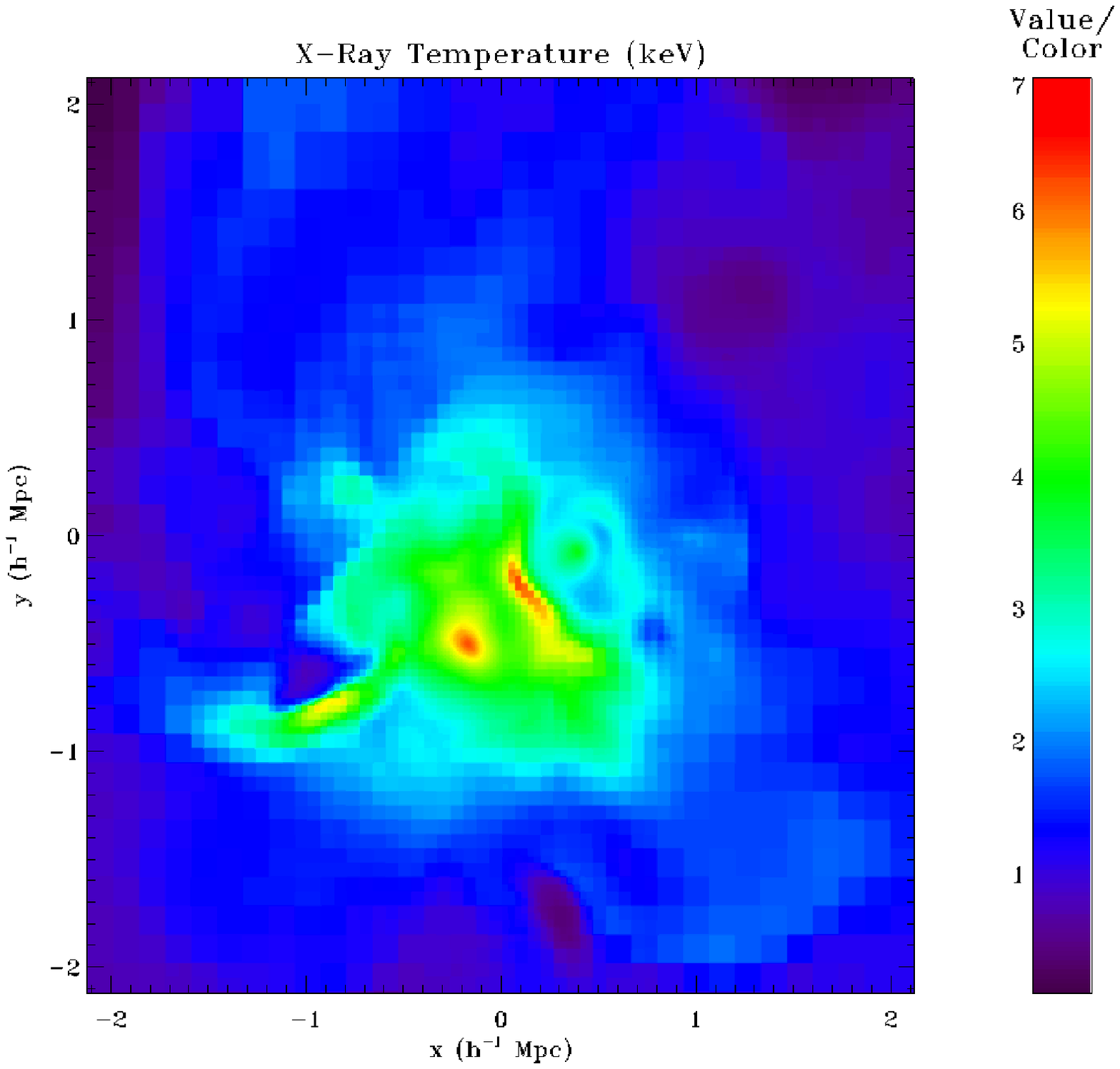, width=8cm}
  \end{center}
\caption{{\em Top:}bolometric X-ray luminosity map. {\em Bottom:} luminosity-weighted temperature map. }  
\label{fauthor-E1_fig:fig4}
\end{figure}

\section{A1750}

 The bimodal shape of A1750 was first observed by EINSTEIN (Forman et al. 1981). The ASCA and ROSAT data (Novicki et al. 1998) apparently showed  evidence for shock heated region between the two components.
A difference of ~1000 km/s in the radial velocity between the two substructures  is a indication of the fact that we are looking at a bound double system (Quintana \& Ramirez 1989).
\begin{figure}[ht]
  \begin{center}
    \epsfig{file=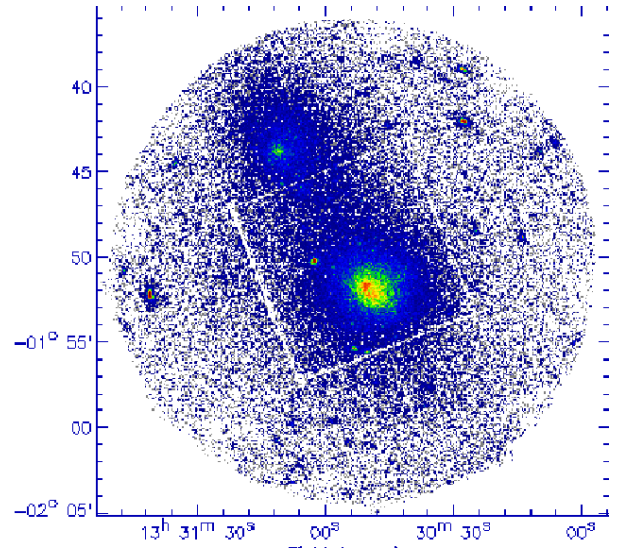, width=8cm}
  \end{center}
\caption{Flux image in the energy band 0.5 - 5.0 keV.}  
\label{fig5:fig5}
\end{figure}

\subsection{Imaging}
The surface brightness profiles are obtained excluding point sources and the other merging component for each subcluster, in the 0.3-1.5 kev image. Both subclusters exhibit an excess in the central region which may suggest the presence of a cooling flow. If this hypothesis is confirmed by further analysis , it will represent significant proof  supporting the suggestion that the components are at a very early stage of the merger because merger events destroy cooling flows.
\begin{figure}[ht]
  \begin{center}
    \epsfig{file=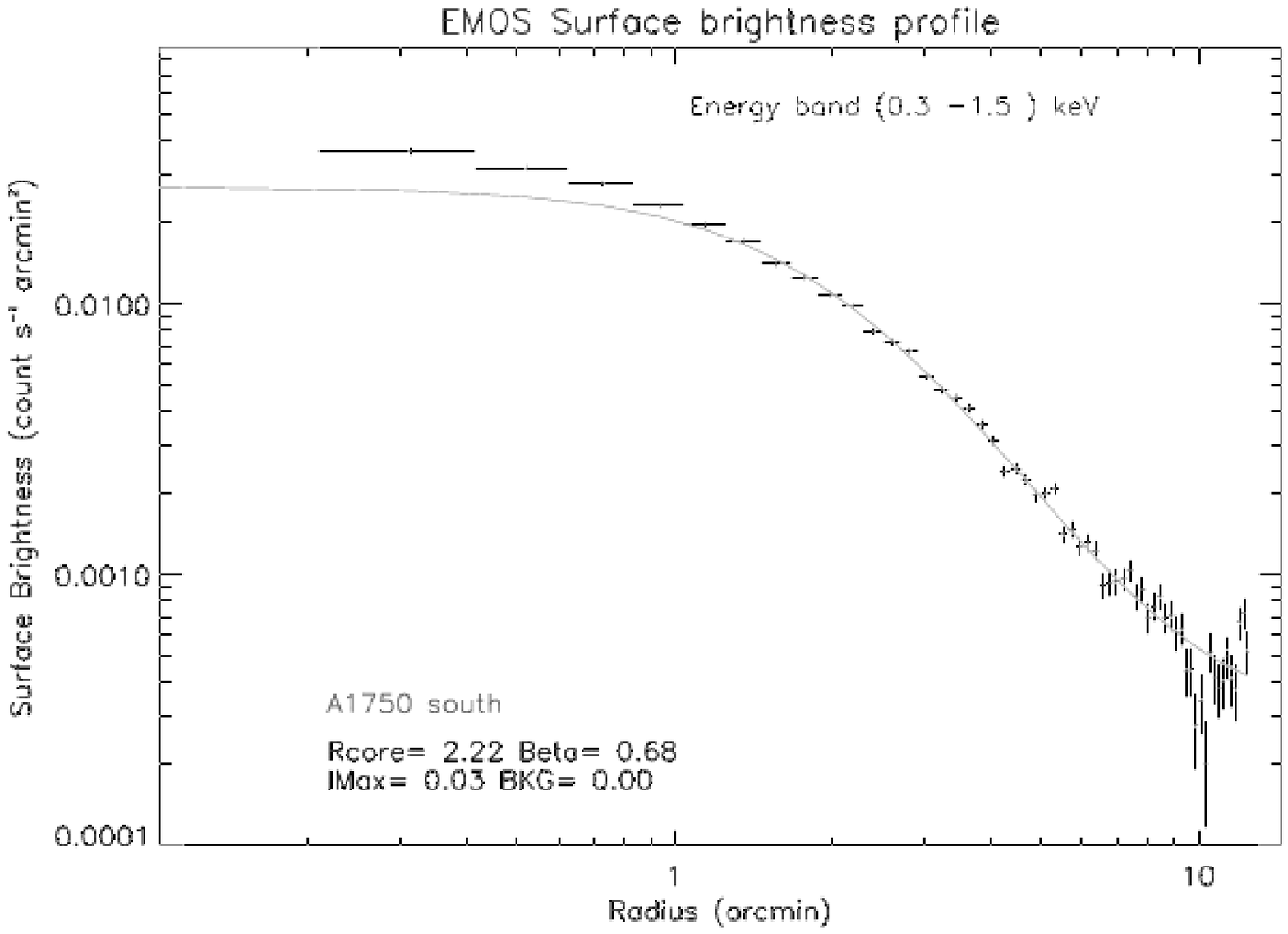, width=7cm}
    \epsfig{file=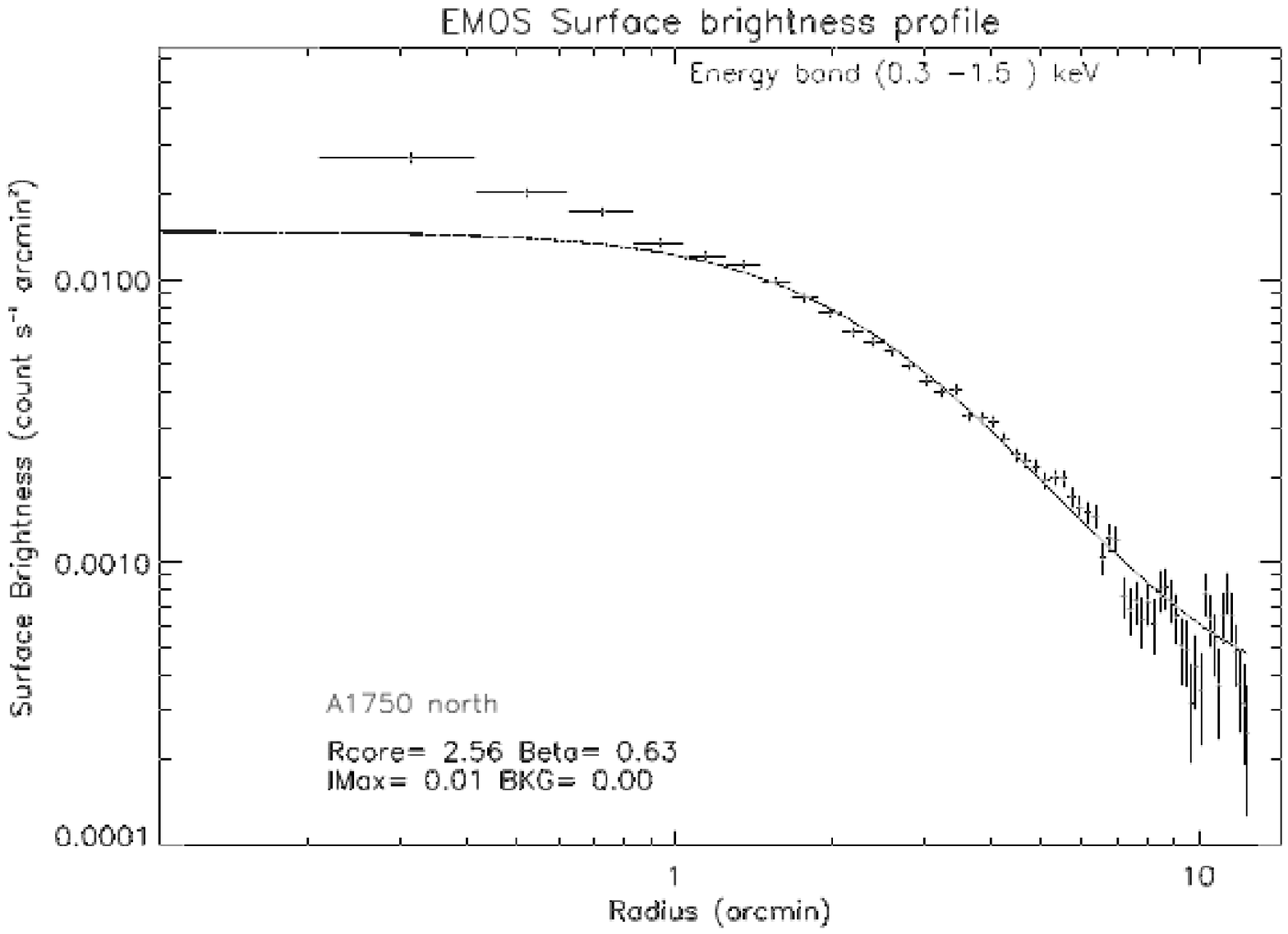, width=7cm}
  \end{center}
\caption{{\em Top:} S-W subcluster surface brightness profile{\em Bottom:} N-E subcluster surface brightness profile }  
\label{fauthor-E1_fig:fig6}
\end{figure}

\subsection{Spectro-imaging}
Temperature map. Contours correspond to the image above after adaptive smoothing 
The temperature map is obtained as for A3921. The A1750 northern subcluster  has a lower temperature (kT~3.3 keV) then the southern one (kT ~ 4.2keV).  The region between the two merging component shows a mean temperature of (5.4 keV with some regions rising to 6.7 keV. This preliminary result needs confirmation from EPN analysis, but taken at face value, it is proof that the two subclusters really are interacting and display a shocked region due to the merger event.

\begin{figure}[ht]
  \begin{center}
    \epsfig{file=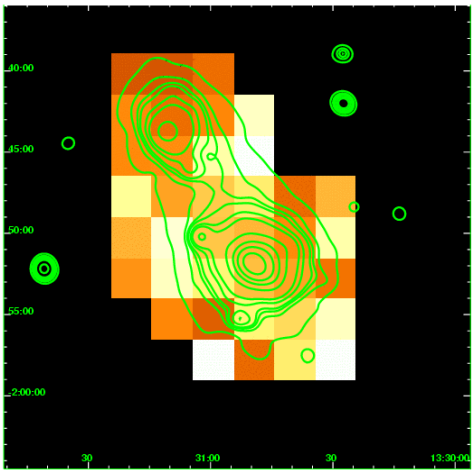, width=8cm}
  \end{center}
\caption{Temperature map. Contours correspond to the flux image in figure \ref{fig5} after adaptative smoothing}  
\label{fauthor-E1_fig:fig7}
\end{figure}

\section{References:}
Belsole E., Sauvageot J-L, Bourdin H., 2001, Sesto 2001 Astrophysical meeting 
Forman et al. 1981, ApJ, 243, L133
Novicki et al. 1998, AAS 193
Quintana \& Ramirez 1989
Sauvageot J-L \& Belsole E. 2001, XXI Moriond Astrophysical Meeting
Teyssier  2001 astro-ph/0111368
\end{document}